\documentclass{article}
\usepackage{amsmath}
\usepackage[dvips]{graphics}
\begin{document}

\title{Metastable states in disordered Ising magnets in mean-field approximation}
\author{P. N. Timonin  \\ %EndAName
Physics Research Institute at Rostov State University,\\344090,
Rostov-on-Don, Russia\\e-mail: timonin@icomm.ru }
\maketitle

\begin{abstract}
The mechanism of appearance of exponentially large number of
metastable states in magnetic phases of disordered Ising magnets
with short-range random exchange is suggested. It is based on the
assumption that transitions into inhomogeneous magnetic phases
results from the condensation of macroscopically large number of
sparse delocalized modes near the localization threshold. The
properties of metastable states in random magnets with zero ground
state magnetization (dilute antiferromagnet, binary spin glass,
dilute ferromagnet with dipole interaction) has been obtained in
framework of this mechanism using variant of mean-field
approximation. The relations between the characteristics of slow
nonequilibrium processes in magnetic phases (such as hysteresis
loop form, thermo-remainent and isothermal remainent
magnetizations, field-cooled and zero-field-cooled thermodynamic
quantities) and thermodynamic parameters of metastable states are
established.
\end{abstract}

\section{Introduction}

The specific feature of disordered magnets is the appearance of large number
of metastable states in magnetic phases. They cause various irreversibility
phenomena in the reaction to the changes of temperature and external
magnetic field, such as the dependence of thermodynamic parameters on the
order and the rate of these changes, the appearance of hysteresis loops,
depending on the field amplitude etc. \cite{1,2,3,4}. These effects
appear to be common to the all types of magnetic disorder - from dilute
magnets with nonmagnetic impurities \cite{2,4} to spin glasses
existing in solid solutions of ferromagnets with antiferromagnets \cite{1,3}%
. This universality of nonequilibrium phenomena compels to suggest
the existence of common mechanism responsible for the appearance
of metastable states in random magnets.

The evidences of the existence of metastable states, the number of
which growth exponentially with the number of sites, has been
obtained in numerical studies of several models of disordered
magnets with short-range exchange \cite{5,6,7,8}. Theoretical
description of the related inergodic phenomena is also possible
mainly with the use of numerical methods, see e. g. \cite{1,4,9}.

The principal analytical results has been obtained in the
Sherrington-Kirkpatrick spin glass model with infinite-range
interaction. It was established that in this model the region of
inergodicity exists in the external field less than the
Almeida-Thouless field $H_{AT}$, \cite{10}, and in the framework
of the replica symmetry breaking scheme by Parisi \cite{11} and
the notion of the hierarchy of macroscopic relaxation times
\cite{12} the thermodynamic parameters of field-cooled and
zero-field-cooled processes has been obtained \cite{1}. But this
is all what is known about the irreversible processes in this
model. In spite of the wide application of such approach to other
mean-field models, see e.g. \cite{13,14}, it is not clear how all
variety of inergodic phenomena resulting from transitions between
metastable states could be described by the methods of
\cite{11,12}.

Meanwhile, theoretical description of irreversible processes could
be made quite simple and apparent if thermodynamic properties of
metastable states, their regions of stability and points of
possible transitions between them would be known at $H<H_{AT}$.
Principally, in the mean-field models this data could be obtained
from (nonaveraged)equations for local magnetic moments. The most
popular example of application of this method in the theory of
disordered magnets is the TAP-equations for local magnetic moments
in the Sherrington-Kirkpatrick spin glass model \cite{15}. It was
established that the number of their solutions is exponentially
large \cite{16}, yet their explicit form has not been determined
by the analytical methods. Now it is still not clear how many of
these solutions correspond to the stable states and how the
divergent barriers between them emerge.

Meanwhile, the results of the investigation of TAP-equations in
case of large but finite interaction range \cite{17} cast some
doubt on the possibility to use the infinite-range models as a
valid approximation for description of real disordered magnets.
The authors of \cite{17} first noticed the consequences of the
qualitative difference of the spectrum of the gaussian orthogonal
ensemble of random exchange in Sherrington-Kirkpatrick model with
all eigenvectors delocalized  and that of short-range random
exchange having localized eigenvectors near its edges \cite{18}.
It was shown in \cite{17} that in $3d$ spin glass with
sufficiently large but finite interaction range, the condensation
of localized modes could drastically change the nature of phase
transition preventing the subsequent condensation of delocalized
modes if localization length index is greater than $2/3$. This
case with the absence of metastable states seems to be not very
common in real disordered magnets but this result demonstrates the
necessity to take into account the qualitative features of
spectral characteristics of short-range random exchange.

The studies of the spectral structure of various short-range
random matrix ensembles have shown that all they have similar
features having localized eigenvectors near the edges and fractal
structure of eigenvectors at the localization threshold
\cite{18,19,20}. Thus one can suppose that universality of the
inergodic phenomena in the different types of disordered magnets
results from the similarity of their spectral properties. The
common mechanism of the appearance of large number of metastable
states in such systems could be an almost simultaneous
condensation of macroscopically large number of almost
nonoverlapping fractal modes near the localization threshold.

Here we must note that the finiteness of the interaction range
does not forbid the quantitative description of such mechanism in
the framework of the mean-field approach. Indeed, the macroscopic
number of condensing modes (order parameter components) greatly
reduce the number of noncondensing ones (order parameter
fluctuations). So we may expect that Ginzburg parameter will be
essentially smaller than that of the homogeneous magnet of the
same dimension for all types of short-range random exchange.

On the basis of such suppositions, we attempt here to develop the
phenomenological mean-field theory of disordered Ising magnets
with the zero ground state magnetization, using some heuristic
suppositions about the structure of the fractal eigenvectors in
random exchange ensembles. In this approach the appearance of
numerous metastable states could be naturally explained and, using
some simplifying assumptions, it is possible to describe
analytically the thermodynamic properties of these states near
transition point. In their turn, these results allow to determine
parameters of all slow irreversible processes and, particularly,
to get the analytical expressions, describing the hysteresis loop
form at arbitrary amplitude of external AC field as well as field
and temperature dependencies of remanent magnetizations in the
considered types of disordered magnets.

\section{Mean-field approximation for disordered Ising magnets}

The Hamiltonian of disordered Ising magnet is
\begin{equation}
\label{1}H=-\frac 12{\sum\limits_{ij}}J_{ij}S_iS_j
\end{equation}
Here $J_{ij}$ is the random exchange matrix, $S_i=\pm 1$. In the majority of
realistic models, $J_{ij}$ are nonzero only for the sites within several
coordination spheres and%
$$
\left| J_{ij}\right| <J_{\max }
$$
Here we consider just such models.

The mean-field approximation for the Hamiltonian in Eq. \ref{1}
consists in the substitution of matrix $J_{ij}$ by the projector
on its largest eigenvalue
\begin{equation}
\label{2}J_{ij}\approx J\sum\limits_{\alpha =1}^{N_0}c_i^\alpha c_j^\alpha
\end{equation}
Here $c_i^\alpha $ are the normalized eigenvectors of $J_{ij}$ corresponding
to the largest (degenerate in case of multisublattice antiferromagnet)
eigenvalue $J$, $\alpha =1,...,N_0$.

Then the Hamiltonian becomes the function of (multicomponent) order parameter%
$$
\eta ^\alpha =N^{-1/2}\stackrel{N}{\sum\limits_i}c_i^\alpha S_i
$$
and one should only calculate the entropy
\begin{equation}
\label{3}S\left( \eta ^\alpha \right) =\ln Tr\delta _{N^{1/2}\eta ^\alpha
,\sum c_i^\alpha S_i}
\end{equation}
to get the inequilibrium thermodynamic potential $$ F\left( \eta
^\alpha \right) =-\frac{NJ}2\sum\limits_{\alpha =1}^{N_0}\left(
\eta ^\alpha \right) ^2-TS\left( \eta ^\alpha \right) $$
Minimization of $F\left( \eta ^\alpha \right) $ would give the
equilibrium values of thermodynamic parameters, corresponding to
the lowest minimum, and those of metastable states, corresponding
to less deep minima. In particular, one can get the average spin
values
\begin{equation}
\label{4}\left\langle S_i\right\rangle_T
=N^{1/2}\sum\limits_{\alpha =1}^{N_0}c_i^\alpha \eta ^\alpha
\end{equation}
In some cases this approximation may give quantitative description of
thermodynamics of the second order transitions in homogeneous magnets
(except for the fluctuation region in the immediate vicinity of transition
point). So it is rather natural to use the mean-field approximation for the
disordered magnets. In this case it seems sufficient to average the results
obtained in the framework of the described above scheme over random $J$ and $%
c_i^\alpha $. Then the thermodynamics of disordered magnets would not
qualitatively differ from that of ideal magnets and, in particular, there
would not appear numerous metastable states in the inhomogeneous magnetic
phases.

The most probable reason for the emergence of exponentially large
number of metastable states lies in the specific structure of
spectrum and eigenvectors of random matrix $J_{ij}$. Indeed, the
eigenvectors of $J_{ij}$ having the described above properties are
localized near the upper (and lower) boundary of the spectrum
\cite{18}. So to describe the transition in random magnet one
should not take in the Eq. \ref{2} the largest eigenvalue but
rather lower one $J$ at the localization threshold, i. e. the
largest eigenvalue from those having delocalized eigenvectors. The
reason is that the macroscopic transition could take place only as
a result of condensation of delocalized mode, while the preceding
condensation of local modes with larger eigenvalues (transition
temperatures) results in the specific transition into Griffiths'
phase, which is not accompanied by noticeable anomalies of
thermodynamic parameters \cite{21,22}. Yet we must note that
according to \cite{17} in case of sufficiently large interaction
range the condensation of local modes could stabilize the
delocalized modes and macroscopic transition would not take place.
So here we assume the range of interaction to be sufficiently
small for macroscopic transition to occur.

Meanwhile, it seems rather probable that in the majority of random
realizations the condensation of just one delocalized mode is not
sufficient for the stabilization of new magnetic phase. For the
dimension $d>2$, delocalized eigenvectors near the localization
threshold has rather sparse (fractal) structure consisting of rare
localization regions connected just by the branching chains. In
other words, there is the set of sites with the structure of
percolation cluster \cite{12}, where $N(c_i^\alpha )^2>>1$, while
on the other sites $N(c_i^\alpha )^2<<1$. In this respect the
modes which are close to the localization threshold differ
essentially from those in the
interior of the spectrum and from the modes of translationally invariant $%
J_{ij}$ which have $N(c_i^\alpha )^2\approx 1$ at almost all
sites.

The evidences of such fractal structure of these nearly localized
modes were obtained in the numerical studies of various ensembles
of short-range random matrices, see \cite{19,20} and references
therein. According to Eq. \ref{4} the condensation of one such
mode results in the appearance of sufficiently large average spins
only on a sparse fractal structure, which would not suffice to
stabilize the modes with lower eigenvalues, being localized, in
general, on the other fractal sets of sites. To be more precise,
the condensation of the nearest to the localization threshold mode
$c_i^0$ can stabilize only those modes with $J_\alpha<J$, which
overlap essentially with it, i. e. having $Nc_i^\alpha c_i^0 >>1$
at almost all sites where $N(c_i^\alpha )^2>>1$. So after the
condensation of the first sparse mode, the second mode having
almost zero overlap with the first one will condense at lower
temperature. Further decreasing of temperature will result in the
condensation of third sparse mode which does not essentially
overlap with the first and second ones and so on. This subsequent
condensation of almost nonoverlapping modes with lower eigenvalues
will take place until sufficiently large average spins appear at
almost all sites. In the intervals between the eigenvalues of
condensing modes there can exist, in general, an arbitrary numbers
of modes which do not condense due to the large overlap with the
previously condensed ones. These modes represents the order
parameter fluctuations and should be omitted in the mean-field
approximation.

Fractal structure of condensing modes suggests that their number
diverges when $N\rightarrow \infty $. Indeed, if the sets of
sites, where the modes considered are mainly localized, have the
fractal dimension $d_f<d$, then the number of the condensing
modes, $N_0$, is of the order $N^{1-d_f/d}$ .

The described mechanism of the transition into inhomogeneous
magnetic phases can rather naturally explain the appearance of
exponentially large number of metastable states. Indeed, the
condensation of one mode in zero field gives rise to two stable
states related by the global spin reversal, while each subsequent
condensation multiplies this number by factor two. Thus there
appears $2^{N_0}\sim\exp(N^{1-d_f/d}\ln2)$ stable states. Spin
configurations in these states will be related by the overturns of
independent groups of spins, corresponding to fractal modes. Just
this structure of the set of ground states has been revealed in
recent numerical study of $3d$ Ising spin glass with $\pm J$
exchange \cite{13}, which makes this mechanism rather probable.

The fact that overlaps of condensing modes almost vanish (i. e.
$Nc_i^\alpha c_i^\beta <<1$ for $\alpha \neq \beta$ for almost all
sites) allows to simplify essentially further considerations. Let
us approximate the corresponding eigenvectors $c_i^\alpha $ by the
set of nonoverlapping (normalized) vectors $e_i^\alpha $,
$e_i^\alpha e_i^\beta =0$ for $\alpha \neq \beta $, which are
equal to $c_i^\alpha $ in the regions where they are mainly
localized ($N(c_i^\alpha )^2>>1$) and zero outside them. Then on
the subspace, spread by the (apparently, orthogonal)
vectors $e_i^\alpha $, $J_{ij\text{ }}$is almost diagonal%
$$
J_{ij}=\sum\limits_{\alpha ,\beta =1}^{N_0}\left( J\delta _{\alpha \beta
}-J_{\alpha \beta }\right) e_i^\alpha e_j^\beta
$$
Here $J_{\alpha \beta }$ is small positively defined matrix,%
$$
\left| J_{\alpha \beta }\right| <<J
$$
Then it is easy to find the mean-field thermodynamic potential, depending on
the multicomponent order parameter%
$$
l_\alpha =N_\alpha ^{-1/2}{\sum\limits_i}e_i^\alpha S_i
$$
$$
N_\alpha ={\sum\limits_i}\theta \left( \left| e_i^\alpha \right| \right)
$$
($\theta \left( x\right) $ is Haviside's step function) and (quasi)local
magnetizations%
$$
m_\alpha =N_\alpha ^{-1}{\sum\limits_i}S_i\theta \left( \left| e_i^\alpha
\right| \right)
$$
It has the form
\begin{equation}
\label{5}F=-\frac 12\sum\limits_{\alpha ,\beta
=1}^{N_0}{\sqrt{N_\alpha N_\beta }}\left( {J\delta _{\alpha \beta
}-J_{\alpha \beta }}\right)
l_\alpha l_\beta -T\sum\limits_{\alpha =1}^{N_0}{N_\alpha S_\alpha \left( {%
l_\alpha ,m_\alpha }\right) }-NHm
\end{equation}
\begin{equation}
\label{6}S_\alpha \left( {l_\alpha ,m_\alpha }\right) =N_\alpha
^{-1}\ln Tr_\alpha \delta _{N_\alpha ^{1/2}l_\alpha ,\sum
{e_i^\alpha S_i}}\delta _{N_\alpha m_\alpha ,\sum {S_i}\theta
\left( \left| {e_i^\alpha }\right| \right) }
\end{equation}
Here $Tr_\alpha $ denotes the sum over spin configuration of those sites
where $e_i^\alpha \neq 0$. According to the above considerations, $N_\alpha
\rightarrow \infty $ in thermodynamic limit,%
$$
\sum\limits_{\alpha =1}^{N_0}N_\alpha =N
$$
and homogeneous magnetization is%
$$
m=\sum\limits_{\alpha =1}^{N_0}\frac{N_\alpha }Nm_\alpha
$$
Thermodynamic potential $F$, Eq. \ref{5}, depends on a small random matrix $%
J_{\alpha \beta }$ and random vectors $e_i^\alpha $. Their form is
determined by the type of the random exchange matrix ensemble. In some cases
it is possible to get some notion on the $e_i^\alpha $ form. For example, in
the spin glass with binary random exchange,%
$$
J_{ij}=\pm J_{\max }
$$
in every bond configuration there are nonfrustrated $d$-dimensional
clusters, i. e. the clusters which have unique spin configuration $\sigma _i$
providing the energy minimum, and%
$$
J_{ij}\sigma _j\approx 2dJ_{\max }\sigma _i
$$
Thus the delocalized eigenvectors with largest eigenvalues could be
approximately constructed via connection of some nonfrustrated $d$%
-dimensional clusters by the branching chains without loops which
are also nonfrustrated at all bond configurations \cite{24}. So
$e_i^\alpha $ can be approximately represented as
\begin{equation}
\label{7}e_i^\alpha =N_\alpha ^{-1/2}\sigma _i^\alpha
\end{equation}
where $\sigma _i^\alpha $ are the spin configuration constructed
as described on the nonfrustrated fractal sets of sites. In dilute
magnets with the concentration of magnetic atoms above the
percolation threshold, the vectors $e_i^\alpha $ can be also
represented in the form Eq. \ref{7} as one can connect by chains
the $d$-dimensional ferromagnetic (antiferromagnetic) clusters
belonging to the percolation cluster. So $\sigma _i^\alpha =1$ in
dilute ferromagnet and $\sigma _i^\alpha =\left( -1\right) ^{{\bf
kr}_i}$ in dilute antiferromagnet.

Yet we must note that in some specific random realization of
matrix $J_{ij}$ its eigenvectors near the localization threshold
and $e_i^\alpha $ could essentially differ from that of Eq.
\ref{7}. Nevertheless we will suppose that this expression
approximates $e_i^\alpha $ reasonably well in the majority of
random realizations so it can be used for the estimates of the
ensemble average of the sums
\begin{equation}
\label{8} u_{n\alpha }=N_\alpha ^{n/2-1}{\sum\limits_i}(e_i^\alpha
)^n
\end{equation}
 As we show below, thermodynamics near the transition point
does not depend on the detailed form of $e_i^\alpha $ being
determined by the several constants $u_{n\alpha }$, Eq. \ref{8}.

The advantage of $F$ representations in the form of Eq. \ref{5} is
the additivity of the entropy. But with arbitrary random matrix
$J_{\alpha \beta }$ this expression is still difficult to analyze.
We can simplify it using rather apparent consideration that
eigenvalues of this matrix, lying in an interval between zero and
some $J_0<<J$, must condense near zero as the farther from
localization threshold the more rare are the modes which do not
overlap with the preceding ones. Then, as the average interval
between eigenvalues is of the order $1/N$, the same order could
have the smallest eigenvalues of $J_{\alpha \beta }$. So we can
approximate $J_{\alpha \beta }$ by the projector on some (random)
vector $r_\alpha $ which properties are determined by the type of
$J_{ij}$ ensemble,
\begin{equation}
\label{9}J_{\alpha \beta }=J_0r_\alpha r_\beta,\;\;\;
\sum\limits_{\alpha =1}^{N_0}r_\alpha ^2=1
\end{equation}
 Here we must note
that the assumption that only one of eigenvalues of $J_{\alpha
\beta }$ is finite, while all others are of the order $1/N$, is
rather rough. It results in merging of the condensation points of
all modes except one and makes the transition to be more sharp,
while actually some modes will condense somewhere between $T=J$
and $T=J-J_0$. Yet the approximation in Eq. \ref{9} allows to
obtain analytical results which agree qualitatively with
experiments, so it could be a starting point for more precise
theory, accounting for distribution of the condensation
temperatures of fractal modes.

Further we will show that the form of $r_\alpha $ could be
determined from the fact that matrix $J_{\alpha \beta }$ defines
the type of the ensemble's ground state.

Then Eq. \ref{5} becomes
\begin{equation}
\label{10} F =  - \frac{J}{2}\sum\limits_{\alpha  = 1}^{N_0 }
{N_\alpha  l_\alpha ^2 }  + \frac{{J_0 }}{2}(\sum\limits_{\alpha =
1}^{N_0 } {\sqrt {N_\alpha  } r_\alpha  l_\alpha ^{} } )^2  -
T\sum\limits_{\alpha  = 1}^{N_0 } {N_\alpha  S_\alpha  \left(
{l_\alpha  ,m_\alpha  } \right)}  - NHm
\end{equation}
Partial entropies $S_\alpha \left( {l_\alpha ,m_\alpha }\right) $
in Eq. \ref{6} can be represented as $$ S_\alpha  \left( {l_\alpha
,m_\alpha  } \right) = \ln 2 - \mathop {\max }\limits_{\varphi
,\psi } \left[ {\varphi m_\alpha   + \psi l_\alpha   - N_\alpha ^{
- 1} \sum\limits_i {\theta \left( {\left| {e_i^\alpha  } \right|}
\right)} \ln ch\left( {\varphi  + \psi e_i^\alpha  \sqrt {N_\alpha
} } \right)} \right] $$ The values ${\varphi _\alpha ,\psi _\alpha
}$ corresponding to the maximum are determined by the equations
\begin{equation}
\label{11} m_\alpha   = N_\alpha ^{ - 1} \sum\limits_i {th\left(
{\varphi _\alpha   + \psi _\alpha  e_i^\alpha  \sqrt {N_\alpha  }
} \right)},\;\;\;
 l_\alpha   = N_\alpha ^{ - 1/2} \sum\limits_i
{e_i^\alpha  th\left( {\varphi _\alpha   + \psi _\alpha e_i^\alpha
\sqrt {N_\alpha  } } \right)}
\end{equation}

Differentiating potential in Eq. \ref{10} over $l_\alpha $ and
$m_\alpha $ we get the equations of state
\begin{equation}
\label{12}T{\varphi _\alpha =H},\;\;\;  J_0 N_\alpha ^{ - 1/2}
r_\alpha  \sum\limits_\beta  {N_\beta ^{1/2} r_\beta  l_\beta  } -
Jl_\alpha   + T\psi _\alpha   = 0
\end{equation}
The stable solutions of Eqs. \ref{11} - \ref{12} corresponding to
the minima of $F$ must have the positively defined Hessian
\begin{equation}
\label{13} G_{\alpha \beta }  = \delta _{\alpha \beta } \left\{
{T\left[ {1 - \sum\limits_i {\left( {e_i^\alpha  } \right)^2 th^2
\left( {\varphi _\alpha   + \psi _\alpha  e_i^\alpha  \sqrt
{N_\alpha  } } \right)} } \right]^{ - 1}  - J} \right\} + J_0
r_\alpha  r_\beta
\end{equation}
For $H=0$ , $T=0$ we have from  Eqs. \ref{11} - \ref{12}%
$$ m_\alpha   = N_\alpha ^{ - 1} \sum\limits_i {sign\left(
{e_i^\alpha  l_\alpha  } \right)} $$
 $$ \left| {l_\alpha  }
\right| = N_\alpha ^{ - 1/2} \sum\limits_i {\left| {e_i^\alpha  }
\right|}
 $$
 Thus there are $2^{N_0}$ stable solutions of  Eqs. \ref{11} - \ref{12}
in this case which differ by the $l_\alpha$ signs . If the
ensemble of random $J_{ij}$ has ground states with $m=0$ in almost
all realizations then we may provide the minimal energy for the
states with $m=0$ putting $\sum\limits_{\alpha  = 1}^{N_0 } {\sqrt
{N_\alpha  } r_\alpha  l_\alpha }=cm$.  There is the unique
function $r_\alpha(e_i^\alpha)$ obeying this condition for
arbitrary $l_\alpha$ signs,
\begin{equation}
\label{14}r_\alpha =\frac{c^{\prime }\sum\limits_isign({e_i^\alpha
})}{\sum\limits_i\left| {e_i^\alpha }\right| }
\end{equation}
Here $c^{\prime }$ is a normalization constant.

Further we consider only disordered magnets with zero
magnetization in ground states such as dilute antiferromagnets,
spin glasses and dilute ferromagnets with dipole interaction
\cite{25}. According to the above considerations on the form of
${e_i^\alpha }$ in dilute magnets and binary spin glasses(see Eq.
\ref{7}), Eq. \ref{14} can be represented as
\begin{equation}
\label{15} r_\alpha   = u_{1\alpha } (N_\alpha /\sum\limits_\beta
{N_\beta  u_{1\beta }^2 } )^{1/2}
\end{equation}

Thus, in the approach outlined, the study of metastable states in
the mentioned above magnets consists in the finding of stable
solutions of  Eqs. \ref{11} - \ref{12}, with $r_\alpha $ in the
form of Eq. \ref{15}. Spin configurations corresponding to the
obtained $l_\alpha $, $m_\alpha $ can be found from the expression
\begin{equation}
\label{16}\left\langle {S_i } \right\rangle _T  =
\sum\limits_\alpha {N_\alpha ^{ - 1/2} e_i^\alpha  \frac{{l_\alpha
- u_{1\alpha } m_\alpha  }}{{1 - u_{1\alpha }^2 }}}  +
\sum\limits_\alpha {N_\alpha ^{ - 1} \theta \left( {\left|
{e_i^\alpha  } \right|} \right)\frac{{m_\alpha   - u_{1\alpha }
l_\alpha  }}{{1 - u_{1\alpha }^2 }}}
\end{equation}
The averaging over disorder reduces to the averaging of solutions
over random $e_i^\alpha $ and $J_0<<J$. Note that the localization
threshold $J$ is not random quantity being the characteristics of
the whole ensemble of random $J_{ij}$.

Smallness of $J_0>0$ means that corresponding distribution
function must have sufficiently narrow bounded support, i. e. the
possible $J_0$ values must be smaller than some $ \overline{J}>0$
obeying the condition $\overline{J}<<J$. Contrary to the case of
the sums of macroscopic numbers of variables $u_{n\alpha}$ in Eq.
\ref{8}, there are no reason to suppose the fluctuations of $J_0$
to be self-averaging, that is, that $\langle
J_0^k\rangle\rightarrow \langle J_0\rangle\ ^k$ when
$N\rightarrow\infty$. So the thermodynamic parameters of
metastable states in the inhomogeneous magnetic phases will not be
the self-averaging quantities being determined by the different
$J_0$ values in different samples. Note that the absence of the
self-averaging property of the stable thermodynamic parameters has
been also observed in the numerical studies of disordered magnets
\cite{1,26}.

\section{Thermodynamics near transition}

In the absence of external field, the equations of state, Eqs.
\ref{11} - \ref{12}, have unique
paramagnetic solution at $T>J$ and a number of stable solutions appears at $%
T<J$. Thus at $T=J$, $H=0$ a transition from paramagnetic phase
into inhomogeneous magnetic one takes place. Let us consider
thermodynamics in the vicinity of this transition which is defined
by the condition
\begin{equation}
\label{17}l_\alpha ,m_\alpha <<1
\end{equation}
In this case the equations for magnetizations of condensing modes,
$m_\alpha$, follow from Eqs. \ref{11} - \ref{12},\ref{15}
\begin{equation}
\label{18} \tau m_\alpha   + \tau _0 u_{1\alpha }^2
mN/\sum\limits_\beta  {N_\beta  u_{1\beta }^2 }  + u_{4\alpha }
m_\alpha ^3 /3u_{1\alpha }^2  = u_{1\alpha }^2 H/J
\end{equation}
and $l_\alpha$ can be expressed via $m_\alpha$,
\begin{equation}
\label{19} u_{1\alpha } l_\alpha   = m_\alpha   + (u_{1\alpha }^2
- 1)H/J + \left( {u_{3\alpha }  - u_{1\alpha } u_{4\alpha } }
\right)m_\alpha ^3 /3u_{1\alpha }^3
\end{equation}
Here $\tau =1-J/T$, $\tau _0=J_0/T$. Hessian, Eq. \ref{13}, has
the form $$ T^{ - 1} G_{\alpha \beta }  = \left( {\tau  +
u_{4\alpha } m_\alpha ^2 /u_{1\alpha }^2 } \right)\delta _{\alpha
\beta }  + \tau _0 r_\alpha  r_\beta $$ From Eqs. \ref{17} -
\ref{18} it follows that $H<<J$, ${\tau <<1}$, $\tau _0<<1$.

It is natural to suppose the sums of macroscopic numbers of
variables $u_{n\alpha}$ in Eq. \ref{8} to be self-averaging
quantities, so we can substitute them by their average values.
Assuming that Eqs. \ref{7} hold for the most disorder
realizations, we get $$ \overline{u_{4\alpha }}  =
1,\;\;\;\overline{u_{3\alpha }}  = \overline{u_{1\alpha }}, $$ Let
us also suppose that $\overline{u_{1\alpha }^2}$ do not depend on
the $\alpha$,
\begin{equation}
\label{20}\overline{u_{1\alpha }^2}=\overline {u_1^2}\equiv
N_0^{-1}\sum\limits_{\alpha =1}^{N_0}u_{1\alpha }^2
\end{equation}
It seems that such approximation could not qualitatively change
results. Giving it up one will just add some fluctuations
to final expressions. Yet it allows to simplify essentially
Eqs. \ref{18} making possible their analytical study.

The constant $\overline{u_1^2}$ can be estimated using Eq.
\ref{7}. From Eqs. \ref{7},\ref{8} it follows $$
\overline{u_1^2}=N_0^{-1}\sum_\alpha\left( \nu _\alpha ^{+}-\nu
_\alpha ^{-}\right) ^2 $$ where $\nu _\alpha ^{+}$ and $\nu
_\alpha ^{-}$ are relative parts of positive and negative values
of $e_i^\alpha $, so $\overline{u_1^2}\leq 1$ . Then
$\overline{u_1^2}=1$ in dilute dipole ferromagnets. In dilute
antiferromagnet the difference $\nu _\alpha ^{+}-\nu _\alpha ^{-}$
can be nonzero only due to uncompensated
spins on the surface of $d$-dimensional antiferromagnetic clusters on which $%
e_i^\alpha $ are mostly localized. Hence $\nu _\alpha ^{+}-\nu
_\alpha ^{-}$ is of the order of the surface to volume ratio of
 $d$-dimensional antiferromagnetic clusters, so $%
\overline{u_1^2}\approx D^{-2}$, where $D$ is the average diameter
(in terms of lattice spacing) of these clusters. Evidently, $D$ is
a function of the concentration of antiferromagnetic atoms, which
goes to infinity when concentration tends to 1.

In the binary spin glass, $\overline{u_1^2}$ depends on the
concentration of ferromagnetic bonds $p$. In this case
$\overline{u_1^2}=1$ for $p>1-p_c$ and $\overline{u_1^2}=D^{-2}$
for $p<p_c$, $p_c$ being the bond percolation
threshold on the lattice of magnetic atoms. At $p_c<p<1-p_c$, $\overline{%
u_1^2}$ dependence on $p$ can be qualitatively described as%
$$ \overline{u_1^2}=\frac{p-p_c+D^{-2}(1-p-p_c)}{1-2p_c} $$

Here we also substitute $N_\alpha$ by their average value
$$N_\alpha=N/N_0.$$
Then, introducing instead of $m_\alpha$ reduced magnetizations $\mu_\alpha $ ,%
$$ \mu_\alpha =m_\alpha/\sqrt{\overline {u_1^2}}$$ and
dimensionless field $$h=\sqrt{\overline{u_1^2}}H/T$$ Eqs. \ref{20}
can be represented as
\begin{equation}
\label{21}\tau \mu_\alpha +\tau _0\mu+\mu_\alpha ^3/3=h
\end{equation}
where $\mu=N_0^{-1}{\sum_\alpha}\mu_\alpha $.

At $\tau >0$ Eq. \ref{21} have unique paramagnetic solution with equal $%
\mu_\alpha =\mu$, which we denote as $\mu_0$. It obeys the
equation
\begin{equation}
\label{22}(\tau+\tau _0)\mu_0+\mu_0^3/3=h
\end{equation}
When $\tau <0$ Eq. \ref{21} have $2^{N_0}-2$ stable inhomogeneous
solutions besides $\mu_0$ which can be represented as $$ \mu
_\alpha   = \sqrt { - \tau } \left( {\sin \varphi  + \sqrt 3
\sigma _\alpha  \cos \varphi } \right). $$  Here
$\sigma_\alpha=\pm 1$ and $\varphi(\tau,\tau_0,h,\Delta)$ is the
solution of the equation
\begin{equation}
\label{23} 3\tau _0 \left( {\sqrt 3 \Delta \cos \varphi  + \sin
\varphi } \right) - 2\tau \sin 3\varphi  = 3h\left| \tau \right|^{
- 1/2},
\end{equation}
\begin{equation}
\label{24}\Delta  = \sum\limits_\alpha  {\sigma _\alpha  /N_0 }
\end{equation}
The parameter $\Delta$, varying in the interval $(-1,1)$, defines
the degree of the inhomogeneity of a metastable state and
$\Delta=\pm 1$ correspond to the paramagnetic state with
$\mu_\alpha =\mu$. All states with equal $\Delta$ have also the
equal magnetizations
\begin{equation}
\label{25} \mu  = \sqrt { - \tau } \left( {\sin \varphi  + \sqrt 3
\Delta \cos \varphi } \right),
\end{equation}
and Edwards-Anderson order parameter
\begin{equation}
\label{26} q = N^{ - 1} \sum\limits_i {\left\langle {S_i }
\right\rangle _T^2 }  - m^2  \approx N_0^{ - 1} \sum\limits_\alpha
{\mu _\alpha ^2 }  - \overline {u_1^2} \mu
^2=3\tau(\Delta^2-1)\cos^2\varphi+(1-\overline {u_1^2})\mu^2
\end{equation}
as well as thermodynamic potential
\begin{equation}
\label{27} {\rm 4}F/TN = (\tau _0  + \overline{u_1^2} \tau )\mu ^2
+ \tau q - 3h\mu  - 4\ln 2
\end{equation}
 These states are stable at $\tau+{\mu _\alpha ^2 }>0$, which is
equivalent to the inequality
\begin{equation}
\label{28}\left| \varphi \right| <\pi /6
\end{equation}

As $\tau _0>0$, the left side of Eq. \ref{23} is a monotonously
growing function of $\varphi $ for $\left| \varphi \right| <\pi
/6$. Hence, there is only one stable solution for $\varphi$ at a
given $\Delta $, which exists in the interval of the fields $$ h_
- < h < h_ +,\;\;\; h_ \pm   = \sqrt 3 h_{AT} \Delta /2 \pm h_c $$
\begin{equation}
\label{29} h_{AT}  = \sqrt { - 3\tau } \tau _0
\end{equation}
\begin{equation}
\label{30} h_c  = \sqrt { - \tau } (\tau _0 /2 - 2\tau /3)
\end{equation}
In this interval the solution of the Eq. \ref{23} can be
approximated by the quadratic function of the field
\begin{equation}
\label{31} \varphi  \approx \frac{\pi }{{12h_c }}\left[ {2h -
\sqrt 3 h_{AT} \Delta  - \frac{{4(2 - \sqrt 3 )\Delta h_{AT} (h_ +
- h)(h - h_ -  )}}{{4h_c^2  - (2 - \sqrt 3 )^2 \Delta ^2 h_{AT}^2
}}} \right],
\end{equation}
which gives the exact values $\varphi(h_{\pm})=\pm\pi/6$ and
$\varphi(\Delta h_{AT})=0.$

From the stability condition, Eq. \ref{28}, and Eqs. \ref{23},
\ref{25} it follows that metastable states are stable in the
region $$ 9(\tau _0\mu -h)^2<-4\tau ^3 $$ which is the band on the
$\mu, h$ plane. The magnetization is a monotonously growing
function of $h$ and $\Delta $ inside this band so the field
dependencies of magnetization can be represented as a set of
uncrossed lines bounded from above and below by the $\mu_0(h)$
curve as shown in Fig. 1.
\begin{figure}
\centering
\includegraphics[85,480][500,640]{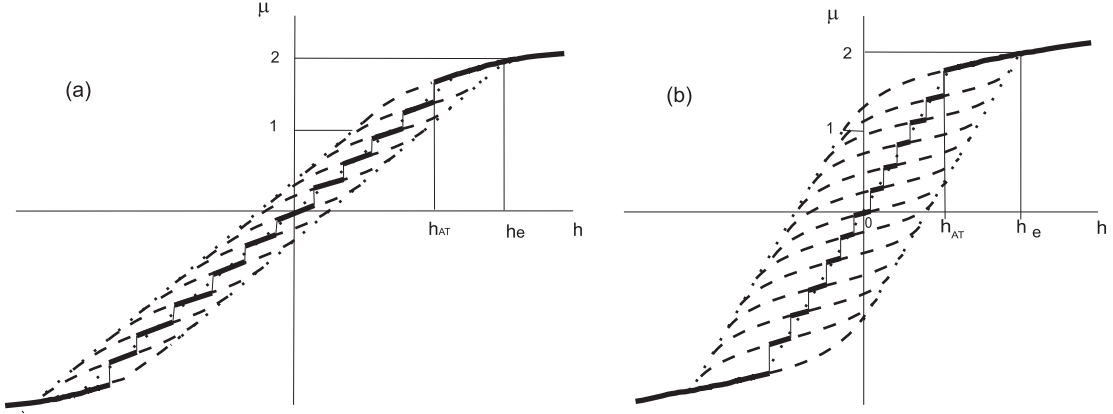}
\caption{Field dependencies of magnetization in metastable states
(dashed lines) and stable states (solid lines) near transition,
(a) - $0<-2\tau<3\tau_0$, (b) - $3\tau_0<-2\tau$.}
\end{figure}

Apparently, the region on this figure, where metastable states
exist, defines the form of hysteresis loop, which appears as a
reaction on a slow AC field with amplitude greater than
\begin{equation}
\label{32}h_e=h_c+ \sqrt 3 h_{AT} /2
\end{equation}
There are certain temperature variations in the loop form as at
$-2\tau<3\tau_0$ only part of metastable states are stable at
$h=0$ and loop is rather slim, see Fig. 1(a), while at
$-2\tau>3\tau_0$ all metastable states are stable at $h=0$ and
loop became more thick, Fig. 1(b). Note also that when amplitude
of AC field is less than $h_e$ the form of hysteresis loop is
defined by the field dependencies of magnetizations in
corresponding metastable states.

Let us also present  the expressions for (dimensionless) magnetic
susceptibility $\chi=\partial\mu/\partial h$,
$$\chi^{-1}=\tau_0-2\tau\cos3\varphi/(\cos\varphi-\sqrt{3}\Delta\sin\varphi),$$
entropy $S$ and heat capacity $C$, $$S=\ln 2-(q+ \overline
{u_1^2}\mu^2)/2 ,$$ $$ C = 1 + \chi \left[ {\frac{3}{2} \cdot
\frac{{\tau _0 \left( {1 - \Delta ^2 } \right)}}{{1 - \sqrt 3
\Delta tg\varphi }} - \tau _0  - \tau } \right] $$

The above expressions allow to get some notion about the field and
temperature dependencies of these quantities. Thus at the
boundaries of stability region, $h =h_\pm$, $q$ and $\chi ^{-1}$
has the lowest values
\begin{equation}
\label{33}q=9\tau \left( \Delta ^2-1\right)/4 -\left(
1-\overline{u_1^2}\right) \tau \left( 3\Delta \pm 1\right)
^2/4,\;\;\;\chi =1/\tau_0,
\end{equation}
while magnetization, entropy and heat capacity are
\begin{equation}
\label{34}\mu=\sqrt{-\tau}\left(
3\Delta \pm 1\right) /2,\;\;\;S=\ln 2+\tau \left( 5\pm 3\Delta \right) /4,%
\;\;\;C=3\left( 1\pm \Delta \right) /2-\tau /\tau _0
\end{equation}
When $|h|$ goes to $h_e$, Eq. \ref{32}, the more homogeneous
states with $\Delta\rightarrow \pm 1$ stay stable and their magnetization tends to $%
\mu_0\left( \pm h_e\right)=\pm 2\sqrt{-\tau}$. However the
limiting values of magnetic susceptibility and heat capacity
differ from those in paramagnetic state: $$\chi _0^{-1}=\tau +\tau
_0+\mu_0^2,\;\;\; C_0=\mu_0^2/\left( \tau +\tau
_0+\mu_0^2\right).$$ In the middle of the stability band (at
$\varphi =0$ or $h=\Delta h_{AT}$) we get:
$$\mu=\Delta\sqrt{-3\tau},\;\;q=-3\tau \left(1-\overline{u_1^2}
\Delta ^2\right),\;\; \chi ^{-1}=\tau_0-2\tau,\;\;S=\ln 2+3\tau
/2,\;\;C=\frac 3{2}\left( 1-\Delta ^2\frac{\tau _0}{\tau _0-2\tau
}\right).$$ In this case with the diminishing of inhomogeneity
when $\Delta\rightarrow\pm1$ or $h\rightarrow \pm h_{AT}$, $\mu$,
$\chi $, $S$ and $C$ tend to the corresponding values of the
paramagnetic phase.

The Almeida-Thouless field $h_{AT}$, Eq. \ref{29}, determines (to the order $%
N_0^{-1}$) the point of the transition into the paramagnetic
phase. To show this let us find the values $\Delta _{eq}$
corresponding to the states with the lowest potential.
Differentiating $F$ , Eq. \ref{27}, over $\Delta $ and
using Eqs. \ref{23}, \ref{25} and Eq. \ref{26} we get%
$$ \frac{{\partial F}}{{\partial \Delta }} =  - NT\sin \varphi
\cos ^3 \varphi ,\;\;\; \frac{{\partial ^2 F}}{{\partial \Delta ^2
}}\left| {_{\varphi  = 0} } \right. > 0 $$ Thus the lowest value
of potential have the states with $\varphi =0$ at given $\tau$ and
$h$ . One can see that Eq. \ref{23} has
solution with $\varphi =0$ when $\Delta =h/h_{AT}$ which is possible at $%
h^2<h_{AT}^2$. When $h^2>h_{AT}^2$ $F\left( \Delta \right) $ has
no minima inside the region $\Delta ^2<1$ in which it is defined
and the minimal values occur at its boundaries for $\Delta
_{eq}=sign(h)$. So the transition into paramagnetic state takes
place at $h=\pm h_{AT}$.

As $\Delta $ is a rational number of the form $2n/N_0-1$ (cf. Eq.
\ref{24})
it can not be exactly equal to $h/h_{AT}$ at all $h^2<h_{AT}^2$. Hence $%
\Delta _{eq}$ is defined so that $\left| \Delta -h/h_{AT}\right| $ is
minimal and can be represented as%
$$
\Delta _{eq}=\sum\limits_{n=1}^{N_0-1}\left( \frac{2n}{N_0}-1\right) \theta
\left( N_0^{-2}-\varepsilon _n^2\right) +sign(h)\theta \left[ h^2-\left( \frac{%
N_0-1}{N_0}\right) h_{AT}^2\right]
$$
$$
\varepsilon _n\equiv \frac h{h_{AT}}-\frac{2n}{N_0}+1
$$
Thus at $h^2<h_{AT}^2$ series of transitions between inhomogeneous magnetic
states takes place at fields%
$$ h_n  = h_{AT} \left( {\frac{{2n + 1}}{{N_0 }} - 1} \right) $$
The value of $\varphi _{eq}$ at $h^2<h_{AT}^2$ corresponding to
$\Delta _{eq}$ is,
see Eq. \ref{23},:%
$$
\varphi _{eq}=\frac{\sqrt 3\tau _0}{\tau _0-2\tau }\sum\limits_{n=1}^{N_0-1}%
\varepsilon _n\theta \left( N_0^{-2}-\varepsilon _n^2\right) $$
Inserting $\Delta _{eq}$ and $\varphi _{eq}$ in the Eqs. \ref{25},
\ref{26} we get the equilibrium values $\mu_{eq}$ and $q_{eq}$ for
$N\rightarrow\infty$:

$$ \mu _{eq}  = \frac{h}{{\tau _0 }}\theta \left( {h_{AT}^2  - h^2
} \right) + \mu _0 \theta \left( {h^2  - h_{AT}^2 } \right) $$ $$
q_{eq}= - 3\tau \left( {1 - h_{}^2 /h_{AT}^2 } \right)\theta
\left( {h_{AT}^2  - h^2 } \right) +\left(
1-\overline{u_1^2}\right) \mu_{eq}^2 $$
Differentiating $m_{eq}$ over $h$ we get the equilibrium susceptibility%
$$ \chi _{eq}  = \tau _0^{ - 1} \theta \left( {h_{AT}^2  - h^2 }
\right) + \left( {\tau  + \tau _0  + \mu _0^2 } \right)^{ - 1}
\theta \left( {h^2  - h_{AT}^2 } \right). $$ The equilibrium
entropy can be obtained by the differentiation of the
equilibrium potential which to the $\varepsilon_n^2$ order is%
$$
F_{eq}=F\left( \Delta =h/h_{AT}\right) -TS_{conf}
$$
where configurational entropy $S_{conf}$ is determined by the logarithm of
the number of states with the same potential $F$,%
$$ S_{conf}=N^{-1}\ln \binom{N_0}{N_0\left( 1-\Delta _{eq}\right)
/2} $$
$S_{conf}$ is of the order $N_0/N$ and can be neglected. Hence at $N_0\rightarrow\infty$%
$$ S_{eq}  = \ln 2 + \frac{{3\tau }}{2}\theta \left( {h_{AT}^2  -
h^2 } \right) - \frac{{\mu _0^2 }}{2}\theta \left( {h^2  -
h_{AT}^2 } \right) $$
For the equilibrium heat capacity we get%
$$ C_{eq}  = \frac{3}{2}\theta \left( {h_{AT}^2  - h^2 } \right) -
\frac{{\mu _0^2 }}{{\tau  + \tau _0  + \mu _0^2 }}\theta \left(
{h^2  - h_{AT}^2 } \right). $$ Let us note that $\mu_{eq}$,
$q_{eq}$ and $S_{eq}$ are continuous at $h^2=h_{AT}^2$, while
$\chi_{eq}$ and $C_{eq}$ undergoes jumps at the transition into
paramagnetic phase.

We must also note that the average equilibrium parameters are
generally unobservable quantities due to the macroscopic free
energy barriers between metastable states. Probably, the
experimental values, which are rather close to them, are obtained
after cooling in small external fields down to $T$ just below $J$
(field-cooled (FC) regime) \cite{1,3} when barriers between
metastable states are relatively small and system could relax into
the lowest (or close to it) state at a sufficiently slow cooling.
In zero field cooled (ZFC) regime when field is applied after
cooling below $T=J$ in zero field, the observed thermodynamic
parameters would differ from equilibrium ones as the system would
at first be trapped in the state with $\Delta =0$ and will stay in
it if applied field does not exceed $h_c$, cf. Eq. \ref{30} and
Fig.1. Thus at $ h<h_c$ the ZFC parameters are those of $\Delta
=0$ metastable states. Their values can be obtained from the
general expressions at $\Delta=0$ and $\varphi=\pi h/6h_c$, see
Eq. \ref{31}. When applied field $h>h_c$, the system relaxes into
the metastable state at the boundary of stability region (on the
lower branch of hysteresis loop) with
$$\Delta_{ZFC}=2(h-h_c)/\sqrt{3}h_{AT}. $$ Inserting this
$\Delta_{ZFC} $ in Eq. \ref{33} and Eq. \ref{34} (with plus sign)
we get the values of thermodynamic parameters the observed
quantities would relax to in ZFC regime at $h_c<h<h_e$. And when
$h>h_e$ ZFC parameters correspond to those of paramagnetic state.
The field and temperature dependencies of some thermodynamic
parameters in FC and ZFC regimes are shown in Fig. 2.
\begin{figure}
\centering
\includegraphics[100,530][530,720]{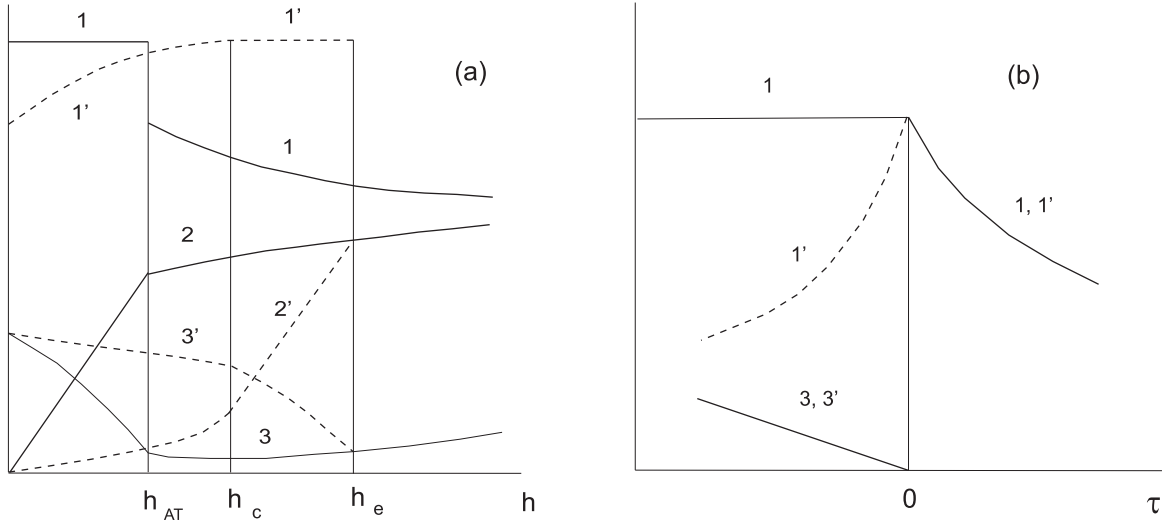}
\caption{Field (a) and temperature (b) dependencies of some
thermodynamic parameters in FC and ZFC regimes, 1-FC
susceptibility, 1'-ZFC susceptibility, 2-FC magnetization, 2'-ZFC
magnetization, 3-FC EA order parameter, 3'-ZFC EA order parameter}
\end{figure}

Similarly, the regions of existence of metastable states, see Fig.
1, and their parameters define the other quantities which are
determined in the slow nonequilibrium processes, such as
thermo-remainent magnetization, $\mu_{TRM}$, which remains after
FC process and subsequent switching off the field, and isothermal
remainent magnetization, $\mu_{IRM}$, remaining after ZFC process
followed by the application for some time (longer than the
intravalley relaxation time) an external field \cite{1,3}.
Apparently, $\mu_{IRM}$ is nonzero only at $ h>h_c$ and an
expression for it can be obtained from Eq. \ref{25} at
 $$\Delta_{IRM}=\min\left[1,2h_c/\sqrt{3}h_{AT},2(h-h_c)/\sqrt{3}h_{AT}\right]$$
 $$
\varphi _{IRM}  \equiv \varphi \left( {h = 0,\Delta _{IRM} }
\right) \approx  - \frac{{\sqrt 3 \pi h_{AT} \Delta _{IRM}
}}{{12h_c }} $$

$\mu_{IRM}$  can also be obtained from Eq. \ref{25} with
$$\Delta_{TRM}=\min\left(1,2h_c/\sqrt{3}h_{AT},h/h_{AT}\right)$$
$$ \varphi _{TRM}  \equiv \varphi \left( {h = 0,\Delta _{TRM} }
\right) \approx  - \frac{{\sqrt 3 \pi h_{AT} \Delta _{TRM}
}}{{12h_c }} $$

The field dependencies of $\mu_{TRM}$ and $\mu_{IRM}$ are shown in
Fig. 3(a).
\begin{figure}
\centering
\includegraphics[60,530][500,730]{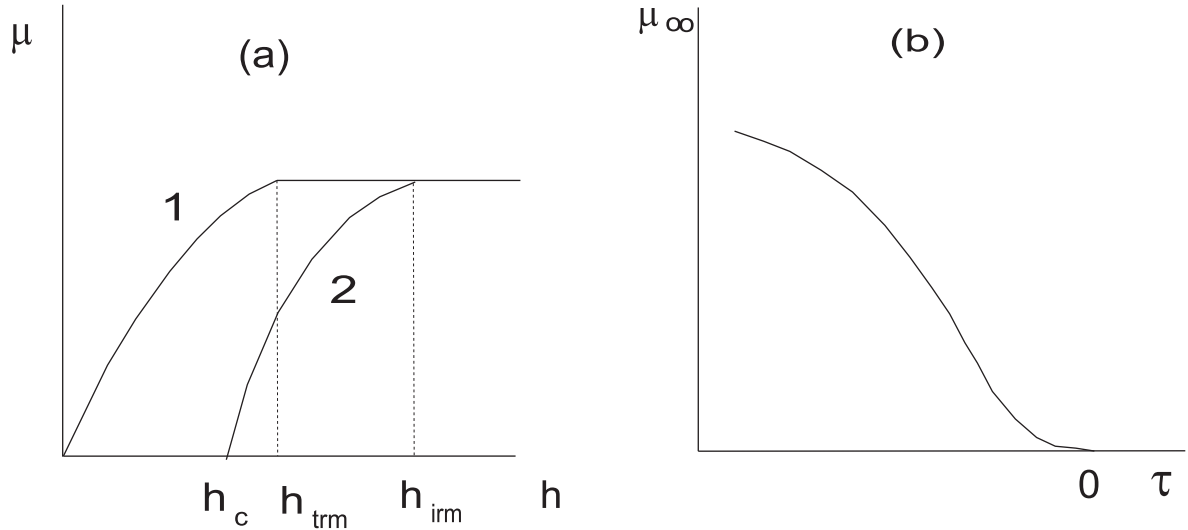}
\caption{(a) - field dependencies of $\mu_{TRM}$ (1) and
$\mu_{IRM}$ (2), (b) - temperature dependence of saturation
magnetization $\mu_\infty$}
\end{figure}
At $h>h_{TRM}=min(2h_c/\sqrt{3},h_{AT})$ $\mu_{TRM}$ becomes
constant, while above $h_{IRM}=min(2h_c,h_e)$ $\mu_{IRM}$ also
saturates at the same value. This value of saturation
magnetization is $$\mu_\infty  = \theta(3\tau _0 + 2\tau)2(-\tau)
^ {3/2}/3\tau _0 + [ - 3(\tau _0 + \tau)]^{1/2} \theta ( - 3\tau_0
- 2\tau )$$ Temperature dependence of $\mu_\infty$ is shown in
Fig. 3(b)

\section{Conclusions}

The main result of the present work consists in the qualitative,
but complete description of properties of all metastable states in
the inergodic phases of random Ising magnets with zero ground
state magnetization and elucidation of their relations to the
parameters of slow irreversible processes. The results depicted in
Fig. 1 allow to describe every conceivable irreversible process
with arbitrary sequence of field and temperature changes.
Qualitative agreement of the obtained here parameters of some such
processes with experiments and numerical studies \cite{1}-
\cite{4} justifies the approximations used in Eqs. \ref{7},
\ref{9}, \ref{15}, \ref{20} and shows that the condensation of
macroscopic number of sparse fractal modes near the localization
threshold do can be a possible mechanism of the appearance of
exponentially large number of metastable states in disordered
Ising magnets. The argument in favor of this mechanism is also the
structure of the set of these states related via overturns of
independent spin groups, corresponding to fractal modes, as just
the same relations between ground states in $3d$ Ising spin glass
with $\pm J$ exchange are revealed in recent numeric studies
\cite{8}.

Let us also note that obtained here results are expressed solely
in terms of statistical characteristics of random exchange matrix.
So the present approach could serve as a starting point for the
developing of more precise quantitative theory of metastable
states in disordered magnets. Such theory should be based on the
detailed studies of the properties of random $J_{ij}$ eigenvectors
near the localization threshold, which we were compelled to
describe here in terms of phenomenological suppositions. One of
the tasks of this theory could be, in particular, the test of the
universality of the properties of the magnets with zero ground
state magnetization, as the results for them differ in present
approach just by the values of $\overline{u_1^2}$.

This work was made under support from Russian Foundation for Basic
Researches, Grant N 98-02-18069.

\end{document}